\pdfoutput=1
\documentclass{article}
\usepackage[round]{natbib}
\usepackage[final]{graphics}
\usepackage{authblk} 
\usepackage{amsmath}
\usepackage[portrait]{geometry}

\begin{document}
\bibliographystyle{abbrvnat}

\title{Computing Curvature for Volume of Fluid Methods using Machine Learning}

\author[1]{Yinghe Qi}
\author[1]{Jiacai Lu}
\author[2]{Ruben Scardovelli}
\author[3]{St\'ephane Zaleski}
\author[1]{Gr\'etar Tryggvason}
\affil[1]{Department of Mechanical Engineering, Johns Hopkins University, MD,  USA}
\affil[2]{Department of Industrial Engineering, University of Bologna, Bologna, Italy}
\affil[3]{Sorbonne Universit\'{e}, CNRS, Institut Jean le Rond d'Alembert, UMR 7190, F-75005, Paris, France}

\maketitle

\begin{abstract}

In spite of considerable progress, computing curvature in Volume of Fluid (VOF) methods continues to be a challenge. The goal is to develop a function or a subroutine that returns the curvature in computational cells containing an interface separating two immiscible fluids, given the volume fraction in the cell and the adjacent cells. Currently, the most accurate approach is to fit a curve (2D), or a surface (3D), matching the volume fractions and finding the curvature by differentiation. Here, a different approach is examined. A synthetic data set, relating curvature to volume fractions, is generated using well-defined shapes where the curvature and volume fractions are easily found and then machine learning is used to fit the data (training). The resulting function is used to find the curvature for shapes not used for the training and implemented into a code to track moving interfaces. The results suggest that using machine learning to generate the relationship is a viable approach that results in reasonably accurate predictions.
\end{abstract}

\section{Introduction}

Computing the evolution of a fluid interface separating immiscible fluids goes back to the beginning of CFD \citep{Harlow:65}. The most common approach is to use one grid for the whole domain and solve the governing equations simultaneously for both fluids using the so-called one-fluid formulation where the different fluids are treated as one fluid with different material properties and a singular source term is added to account for surface forces.  The use of the one-fluid form of the governing equations requires an index or marker function to identify the different fluid and several methods are available to advect the index function \citep{TSZ:2011}. Of those, the Volume of Fluid (VOF) approach, where volume fraction of one fluid is used as an index function, is one the oldest \citep{Hirt:81} and most widely used strategy. 

One of the biggest challenge in VOF methods is the computation of surfaces forces. Major progress was made by the introduction of the Continuous Force method \citep{Brackbill:92}  which made it possible to compute the surface force at fluid interfaces in a reasonably reliable way, by numerically differentiating an normal vector field extended off the interface and using the geometrical identity $\kappa=\nabla \cdot {\bf n}$. A similar strategy was also used by \citet{Lafaurie:94} who worked directly with the stress field. Those approaches, coupled with the Piecewise Linear Interface Calculation (PLIC) strategy \citep{Youngs:82} to advect the volume fraction,  played a major role in making the VOF method  one of the most---if not the most---used method for simulations of flows with sharp, moving interfaces. More recent methods have improved on the above technique by using 1) the so called ``balanced-force methods'' (defined in \citet{Francoisetal2006} but used previously by \citet{RenardyRenardy2002}; and 2) the so-called Height Functions method that allow a point on the interface to be found with fourth order accuracy \citep{Borniaetal2011}. The Height-Function method yields precise results in 2D \citep{Popinet2009}, decreasing spurious currents to machine accuracy, and somewhat less precise results in 3D \citep{Owkesetal2015}. Despite this success, the Height-Function method is limited to situations where the number of grid points per radius of curvature is about ten or more. For smaller radii, one needs to resort to various kinds of fitting. One either directly fits the area or volume under a curve to the volume fraction \citep{RenardyRenardy2002} or finds approximate points on the interface and fits a paraboloid curve or surface the set of points. More complex methods have been suggested \citep{Owkesetal2015}. Even for these more advanced methods, when there are less than ten points in the radius of curvature the error remains large, of the order of one. A review of recent numerical methods for surface tension may be found in \citet{Popinet2018}.

\begin{figure}
\centerline{\scalebox{0.6}{\includegraphics{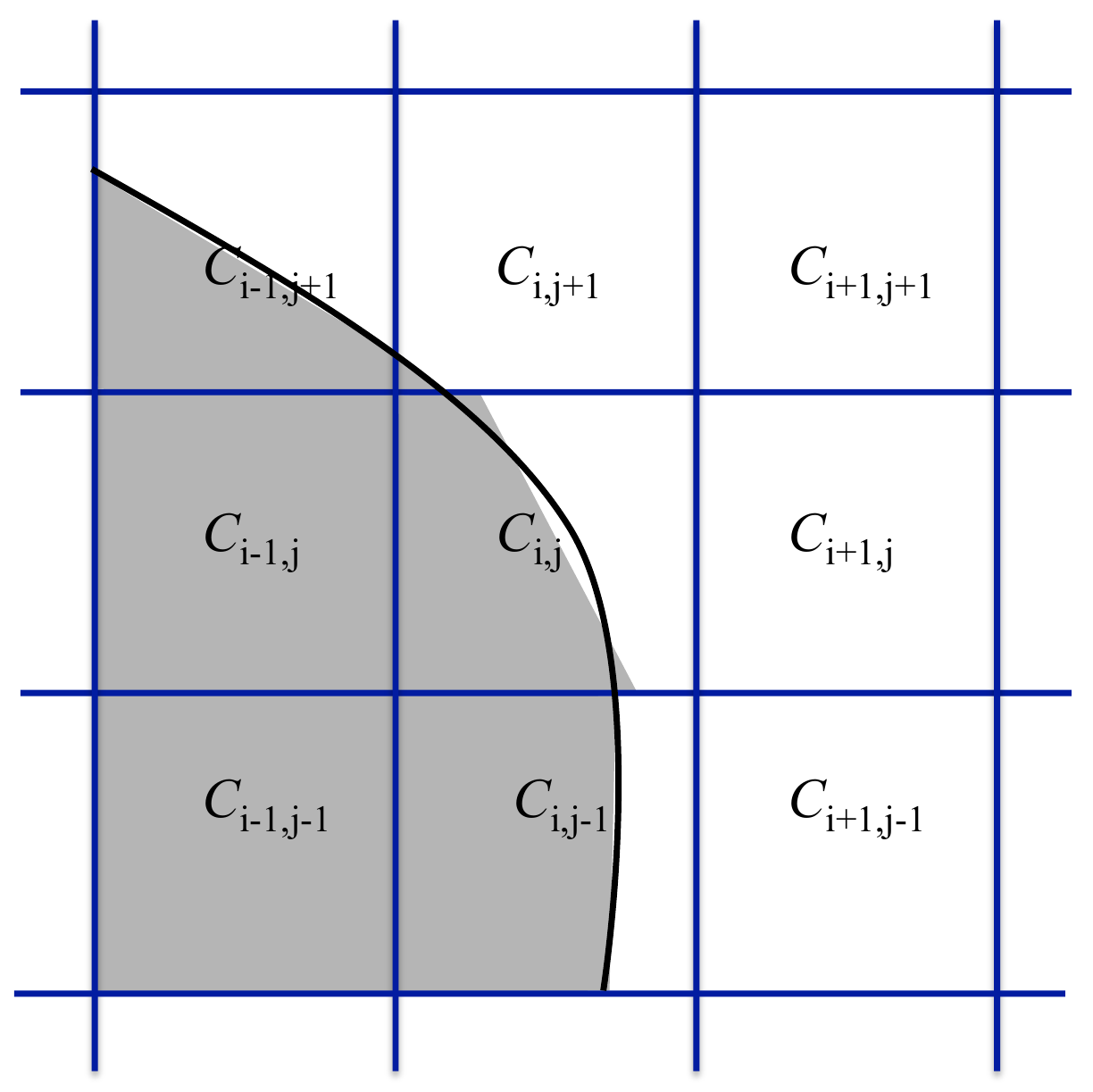}  } }
\caption{Sketch showing an interface crossing several computational cells and the volume fractions constructed by the PLIC method.}
 \label{Sketch}
\end{figure}

Figure \ref{Sketch} shows an interface cutting though several cells and the void fraction in each cell. Although we show the original interface as a line in the figure, in actual computations the only information available is the volume fraction, $C_{i,j}$, in the cell that we are considering and the adjacent cells. We focus on two-dimensional flow below, to simplify the discussion, but the approach should carry over to fully 3D flows in a straightforward way. To find the curvature in each interface cell, $\kappa_{i,j}$, we seek a functional relationship
\begin{equation}
h \kappa_{i,j}=f \Biggl( 
\begin{bmatrix}
    C_{i-1,j+1}& C_{i,j+1} &C_{i+1,j+1} \\
    C_{i-1,j}& C_{i,j}  &C_{i+1,j} \\
    C_{i-1,j-1}& C_{i,j-1} &C_{i+1,j-1} \\
\end{bmatrix}
\Biggr) 
\label{equation1}
\end{equation}
relating the curvature to the nine volume fractions in and around the cell denoted by $(i,j)$. Since the curvature has dimension one over length, it is multiplied by the cell width $h$ (assuming that the cell height and width are the same), to make the expression nondimensional. 

Instead of attempting to derive the relationship expressed by equation (1) analytically, here we examine an alternative approach and find the relation between curvature and the volume fractions using Machine Learning. The closest approach to ours was developed by \citet{Meieretal2002}, where the curvature is found by fitting a dataset generated using circles of different sizes. However, instead of using the nine volume fractions in equation (\ref{equation1}), \citet{Meieretal2002} use three variables accounting for the volume fraction in the nine cells, the volume fraction in the center cell, and the ``tilt'' of the interface.
\begin{figure}
\centerline{\scalebox{0.6}{\includegraphics{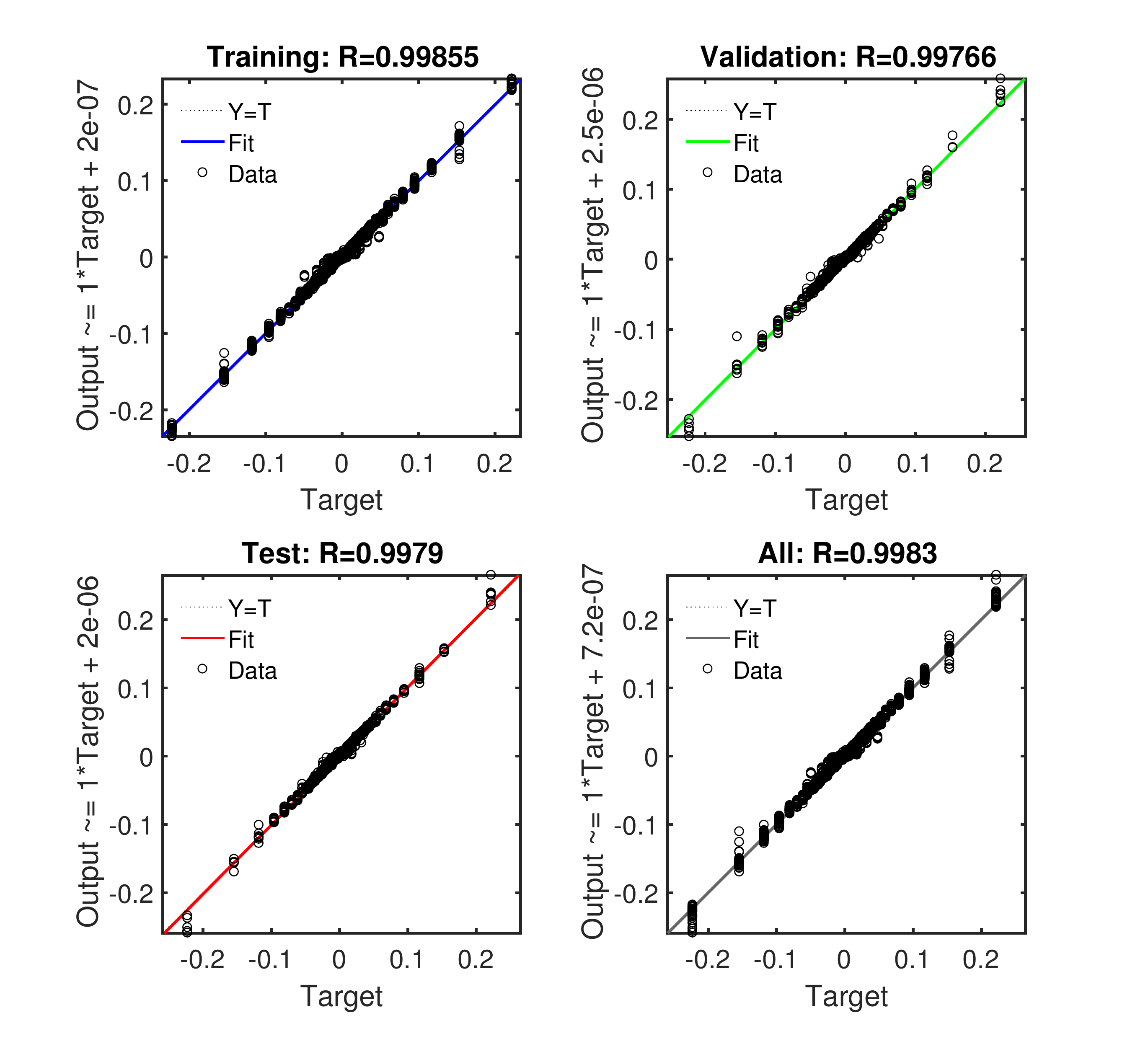}  } }
\caption{Plots of the fitted curvature  versus the exact curvature. (a) training data; (b) test data; (c) validation data; (d) complete dataset. Plots generated by the Matlab Neural Network Toolbox.}
 \label{Statistics1}
\end{figure}

\section{Learning the Curvature and Testing}

Finding the functional relationship between curvature and volume fractions---as expressed by equation (\ref{equation1})---using Machine Learning, involves three steps:

\begin{itemize}
\item Generating of a synthetic dataset, spanning a large range of interface curvatures and orientations with respect to the grid.  
\item Fitting the data using Neural Networks (the learning stage) to find the curvature as a function of the volume fractions.
\item Testing the fit using shapes not used for the fitting
\end{itemize}

To generate a data set containing curvature and volume fractions for well-defined shapes where the curvature and volume fractions are easily found, we use circles of varying sizes (as in \citet{Meieretal2002}). The circles are placed on a grid, and the volume fraction in each cell found. Since we work with the curvature scaled by the grid size, one grid is sufficient. The volume fraction in each grid cell is found by integrating the area under a circle crossing the cell and for cells away from the interface the volume fraction is either zero or one and can be determined by checking if the distance from the center of the cell to the center of the circle is shorter or longer than the radius of the circle.  For each circle we gather data for both positive and negative curvatures, since the sign is defined by whether the fluid inside or outside the circle is identified by $C=1$. By using circles we first of all ensure that the single curvature for each circle is known exactly and we do not have to deal with the question of exactly where in a cell it is computed, and secondly, our data is completely symmetric with respect to rotations and reflections. For the results presented here our data set is generated using a $1 \times 1$ domain resolved by a fixed grid of $2000 \times 2000$ grid points and circles of different radii with a center in the middle of the grid. The radius of the circles ranges from 0.00225 to 0.475 (so that there are 9 grid points across the smallest circle). The scaled curvature ($h \kappa$ thus ranges from 0.001 to 0.222. A total of 65 circles are used, going in increments of 0.001 between the smallest one and the one with radii of 0.05, and then in increments of 0.025 for the rest of the range. This results in 188,232 rows of data, each relating the curvature and the nine void fractions.
\begin{figure}
\centerline{\scalebox{0.6}{\includegraphics{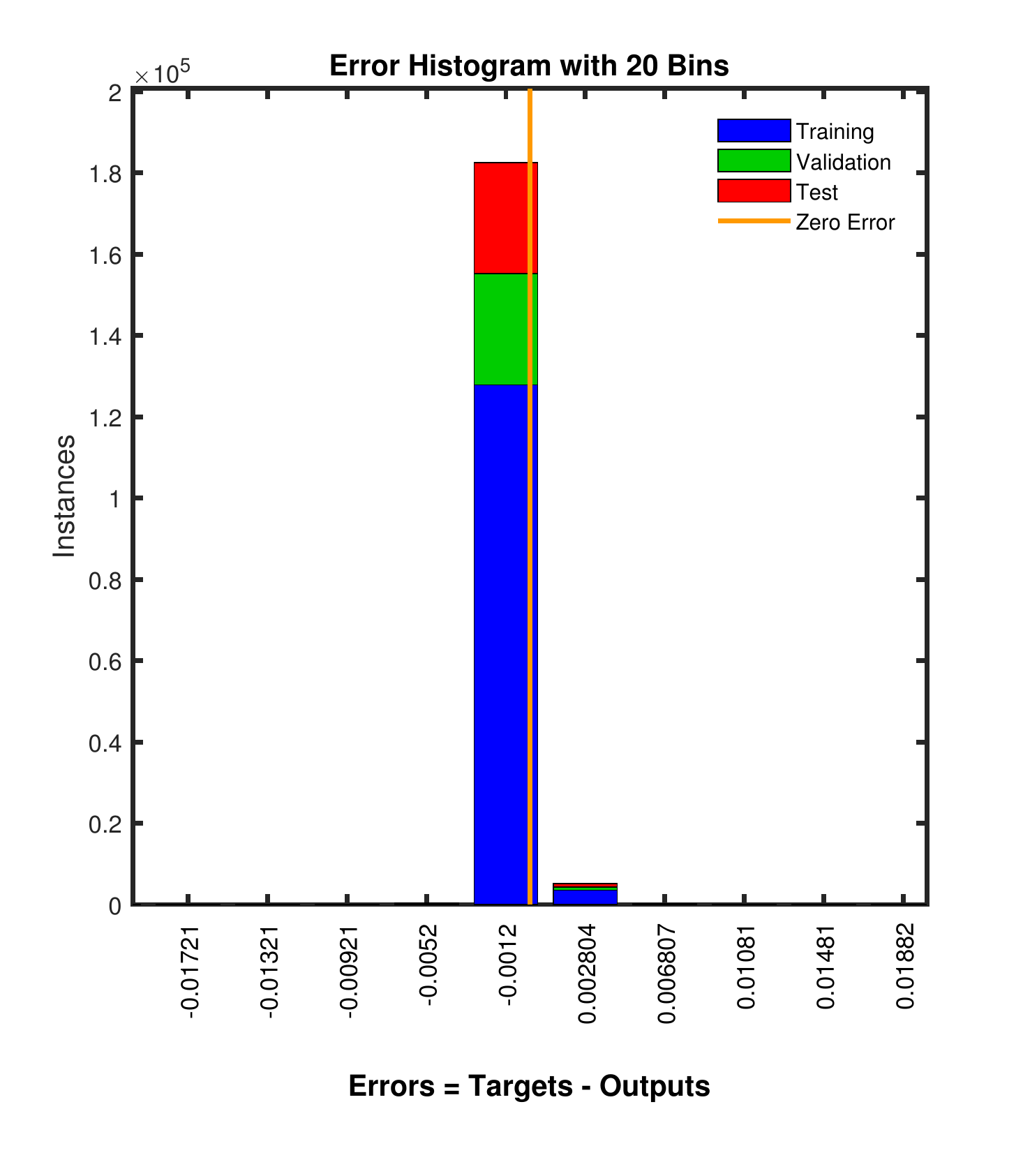}  } }
\caption{The distribution of the error for the fit shown in figure 2. Plots generated by the Matlab Neural Network Toolbox.}
 \label{Statistics2}
\end{figure}

The fitting is done using Matlab and the built-in Neural Network Toolbox. The curvatures constitute a $1 \times N_D$ array of "targets" and the void fractions a $9 \times N_D$ array of inputs, where $N_D$ is the total number of relationships (188,232 for the dataset used here).
After the data has been loaded, the Matlab function $train$ is used to relate the curvature to the void fractions.  We use the default settings of a two-layer feed-forward network with 100 hidden neurons and linear output neurons using the Levenberg-Marquardt process for the back propagation. The data is divided in three parts for training, testing and validation, and we use the default of  70\%, 15\% and 15\%, respectively. The training stops when the mean square error of the validation sample stops improving or the maximum number of steps (1000 in our case) is reached. At that point the network is saved as a Matrix-Only Function. Figure \ref{Statistics1} shows the quality of the fit. The curvature as found by the neural network is plotted versus the exact curvature for each point in the dataset. If the fit was perfect, all the points should lie on the 45 degree line. In (a) fit is shown for the training data; (b) shows the same thing for the test data; In (c) we examine the validation data; and in (d) the fitted curvature and the exact curvatures are compared for the complete dataset. The solid line is a least squares fit through the points shown in each frame.  Figure \ref{Statistics2} shows the error distribution for the final fit. Obviously, most of the training data points have close to zero error, but even the validation points are closely distributed around the origin. 

\begin{figure}
\centerline{ {\scalebox{0.6}{\includegraphics{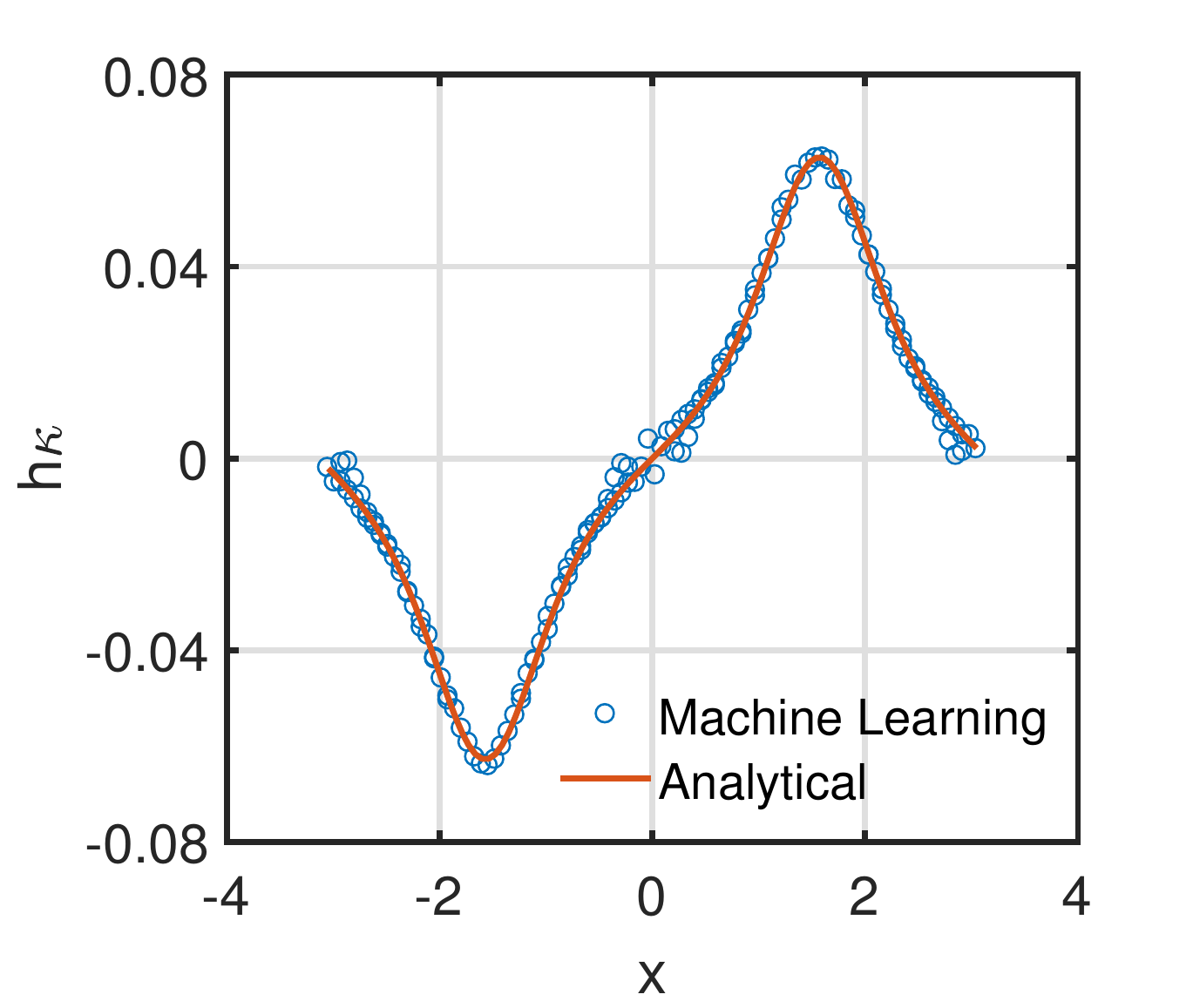}  } }  {\scalebox{0.6}{\includegraphics{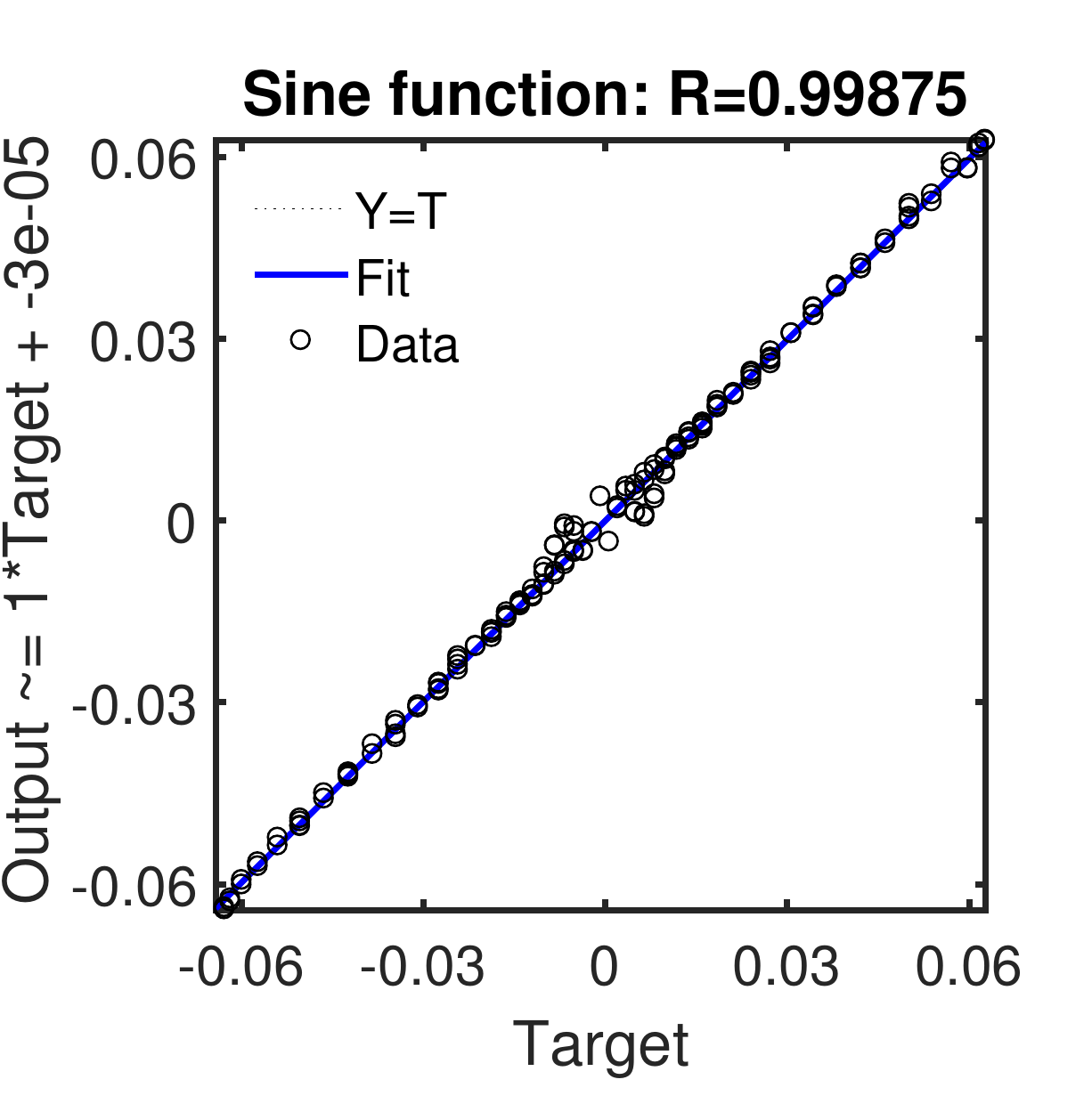}  } }  }
\caption{(a) Plots of the curvature  found by machine learning (circles) and the exact curvature (line) for a sine wave. (b) The curvature  found by machine learning versus the exact curvature.}
 \label{Statistics3}
\end{figure}

The fitting (or learning) results in a function that is saved and can be called from other programs to find the curvature for values of the volume fractions resulting from arbitrarily shaped interfaces. 
To test the fit on a more complex curve, where the curvature varies along the interface, we look at a large amplitude sine wave, defined by $y(x)=A \sin (x)$. The curvature can be found analytically: $\kappa(x) = {-A \sin (x)  \times ( 1+ A^2 \cos^2(x))^{-3/2} }$.
For the results presented here $A=1.0$. The curve divides a rectangular $ 2 \pi \times 2 \pi$ domain into two regions, and the value of the volume fraction of an index function is set to unity in cells below the curve and zero above the curve. The domain is resolved by a $100 \times 100$ grid and the volume fractions in cells crossed by the curve are constructed in the same way as for the learning cases, by representing the curve by connected marker points and finding the fractional area on one side of the curve.  The curvature was calculated analytically for the midpoint of each cell. 
In figure \ref{Statistics3}(a) we plot the fitted curvature and the exact curvature versus the $x$ coordinate, for one value of the amplitude. Overall the agreement is reasonably good, although the points near zero curvature show some scatter. The quality of the fit is also shown in figure \ref{Statistics3}(b), where the curvature from the neural network is plotted versus the exact curvature. If the agreement was perfect, all the points should lie on the 45-degree line. We have also tested the fit using larger amplitudes for the sine function and different grid resolution and find similar results.
As an aside we mention that initially our data set lacked points for very small curvatures (nearly flat interfaces) and in those cases we generally saw large errors where the curvature was small. A more complete data set solved this problem, emphasizing the need for the data to span all possible cases.

\begin{figure}
\centerline{ {\scalebox{0.45}{\includegraphics{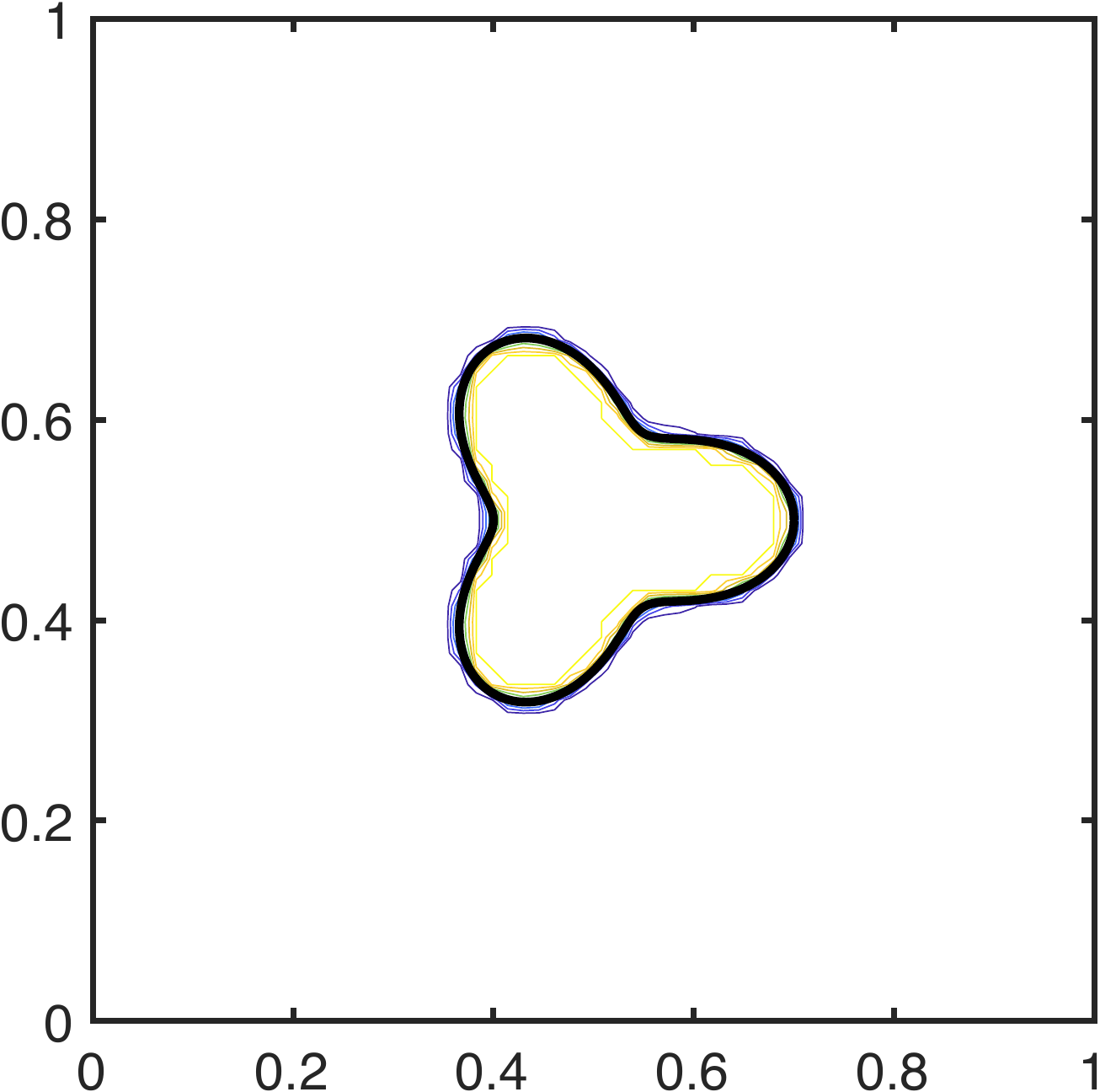}  } }  {\scalebox{0.55}{\includegraphics{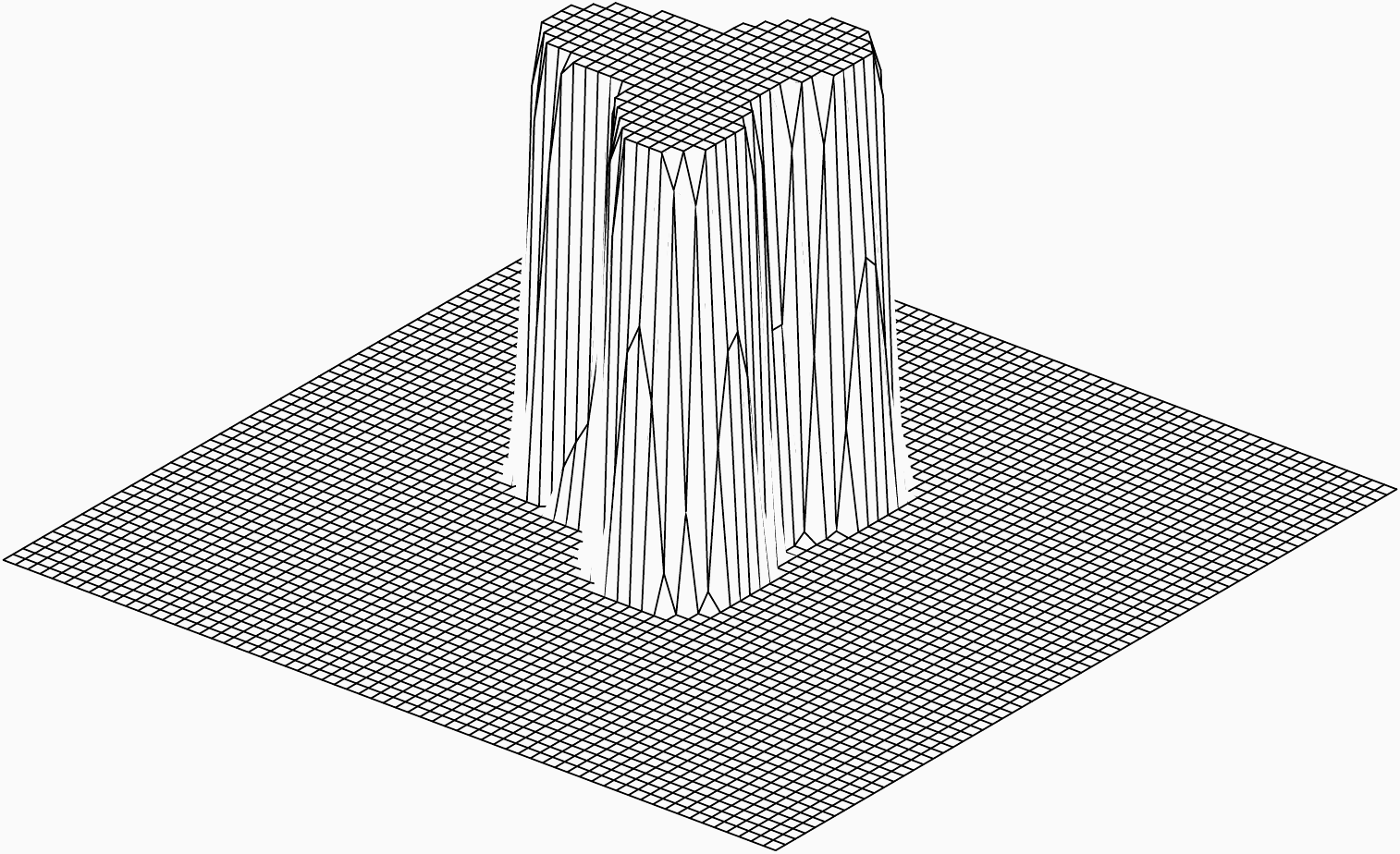}  } }  }
\caption{The initial conditions used to test the curvature routine generated by the neural network.}
 \label{Implemented1}
\end{figure}

\section{Implementation in a Flow Solver}

To test how finding the curvature by machine learning works in a flow solver, we have implemented the function developed by the neural network machine learning in a finite-volume/front-tracking code, where the interface is tracked by connected marker particles advected by the flow. The flow solver is a standard finite volume scheme implemented on a regular staggered grid, where all spatial variables are approximated by centered second-order finite differences and the time integration is done by a third-order Runga-Kutta method \citep{TSZ:2011}. A simple SOR iteration is used to solve the pressure equation. All boundary conditions are taken to be no-slip walls.The interface is represented by connected marker points and the volume fraction (and thus the index function) is found directly from the front \citep {Aulisa:03, Aulisa:04}, in the same way as for the learning data and the sine function example above.  To find the surface force we follow a procedure similar to \citet{Meieretal2002} and first find the normal to the interface by differentiating the volume fraction field, and then use it and the volume fraction to find a straight line approximation, $\Delta l$,  to the length of the interface in each cell. The surface force per unit volume is then given by $ {\bf f}^{\sigma}_{i,j} = \sigma \kappa_{i,j} {\bf n}_{i,j} \Delta l_{i,j} / h^2$. The surface tension is computed in each pressure cell and averaged to find the value at the velocity nodes. We note that for interface cells we divide the surface force by the average density of the different fluids, following \citet{Meieretal2002}, instead of the actual density in the cell. The interface is smoothed slightly at each time step to suppress small ``wiggles.'' 

\begin{figure}
\centerline{ {\scalebox{0.5}{\includegraphics{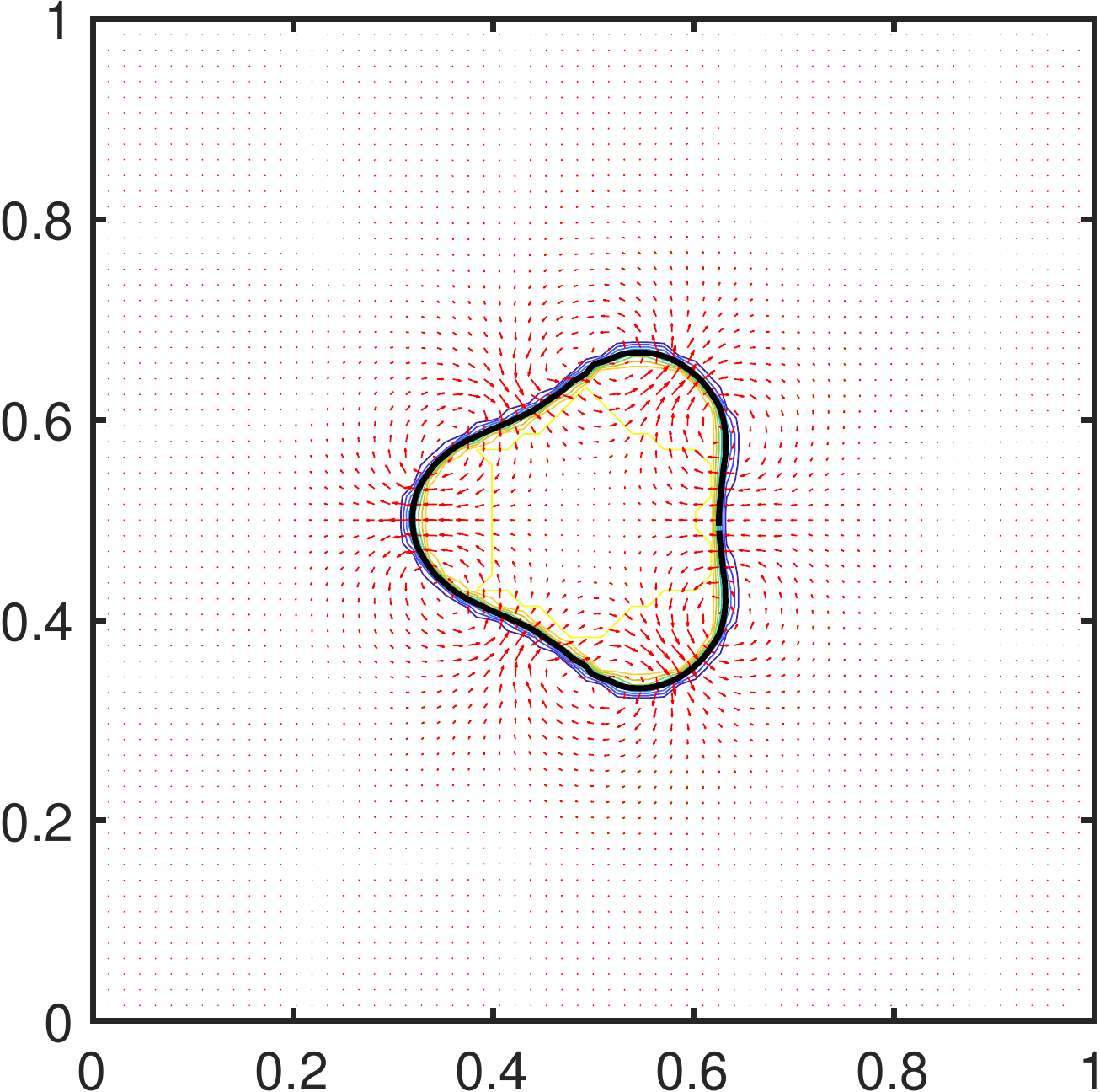}  } }  {\scalebox{0.32}{\includegraphics{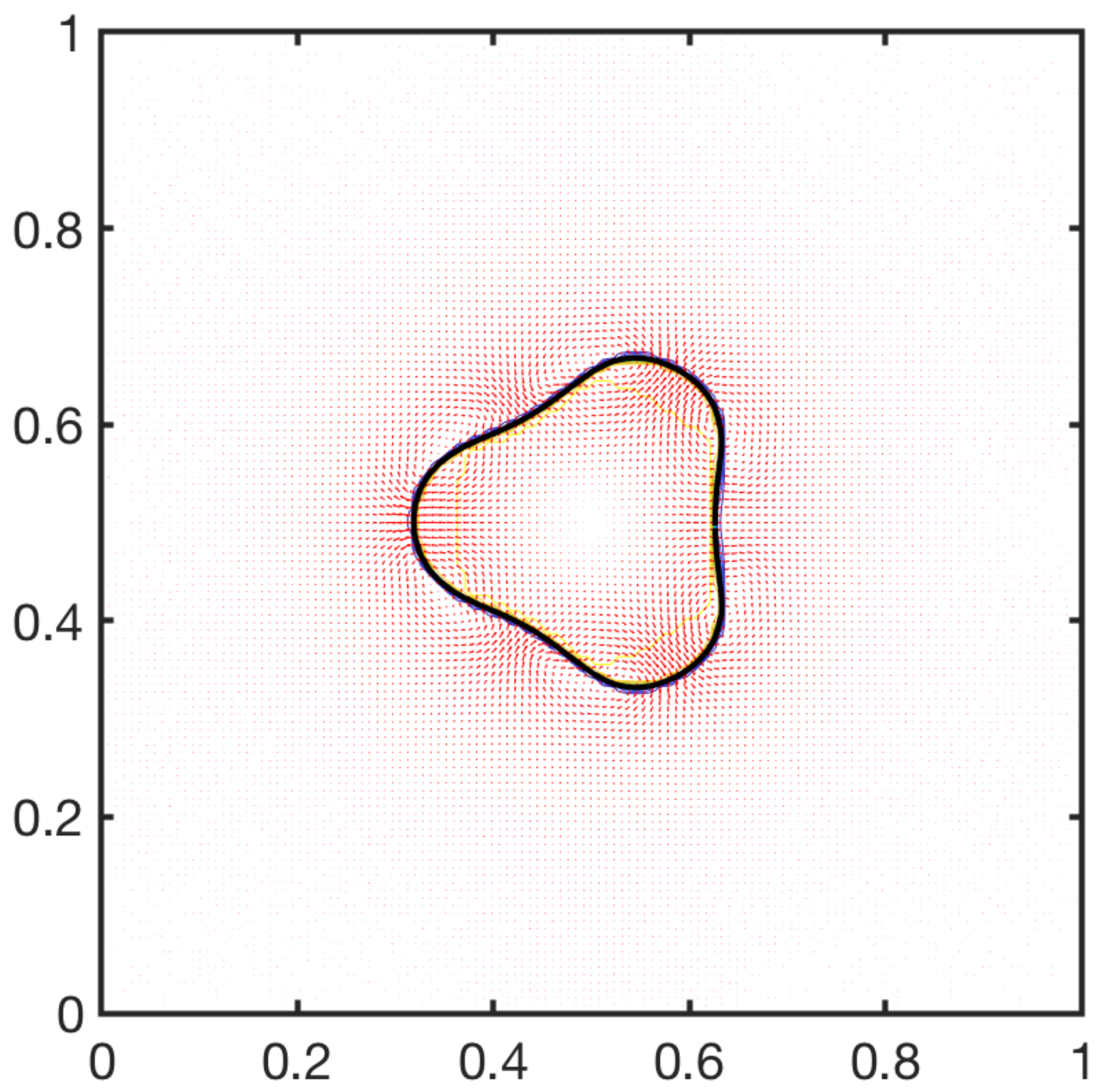}  } }  }
\caption{The interface shape and the velocity field at time 0.02 as computed on a $64^2$ grid (left) and a $128^2$ grid (right).}
 \label{Implemented2}
\end{figure}

Figure \ref{Implemented1} shows the initial conditions for one case. Here the initial interface is a three-armed ``star'' specified by $r(\theta)=R_o+A_o \times \cos(n \theta)$ with $n=3$, $R_o = 0.15$, and $A_o = 0.05$, defining a drop with density $\rho_d = 2.0$ and viscosity $\mu_d=0.02$. The ambient fluid has $\rho_o = 1.0$, viscosity $\mu_o=0.01$ and the surface tension is $\sigma=10$. We follow the motion of the drop  as surface tension pulls it into a circular shape. In figure \ref{Implemented1} the interface and contours of the index function are shown on the left and a 3D view of the index function on the left. The fluid solver uses a $64^2$ grid to resolved the $1 \times 1$ domain and the time step is equal to $\Delta t = 0.005$. 
The interface shape, contours of the index function and the velocity field at one time are shown  in figure \ref{Implemented2} for two different grid resolutions.  At the time shown the shape been ``inverted'' in the sense that ``valleys'' have replaced ``bulges'' and vice versa, although the amplitude is still growing, as seen in the velocity field. 

\begin{figure}
\centerline{ {\scalebox{0.35}{\includegraphics{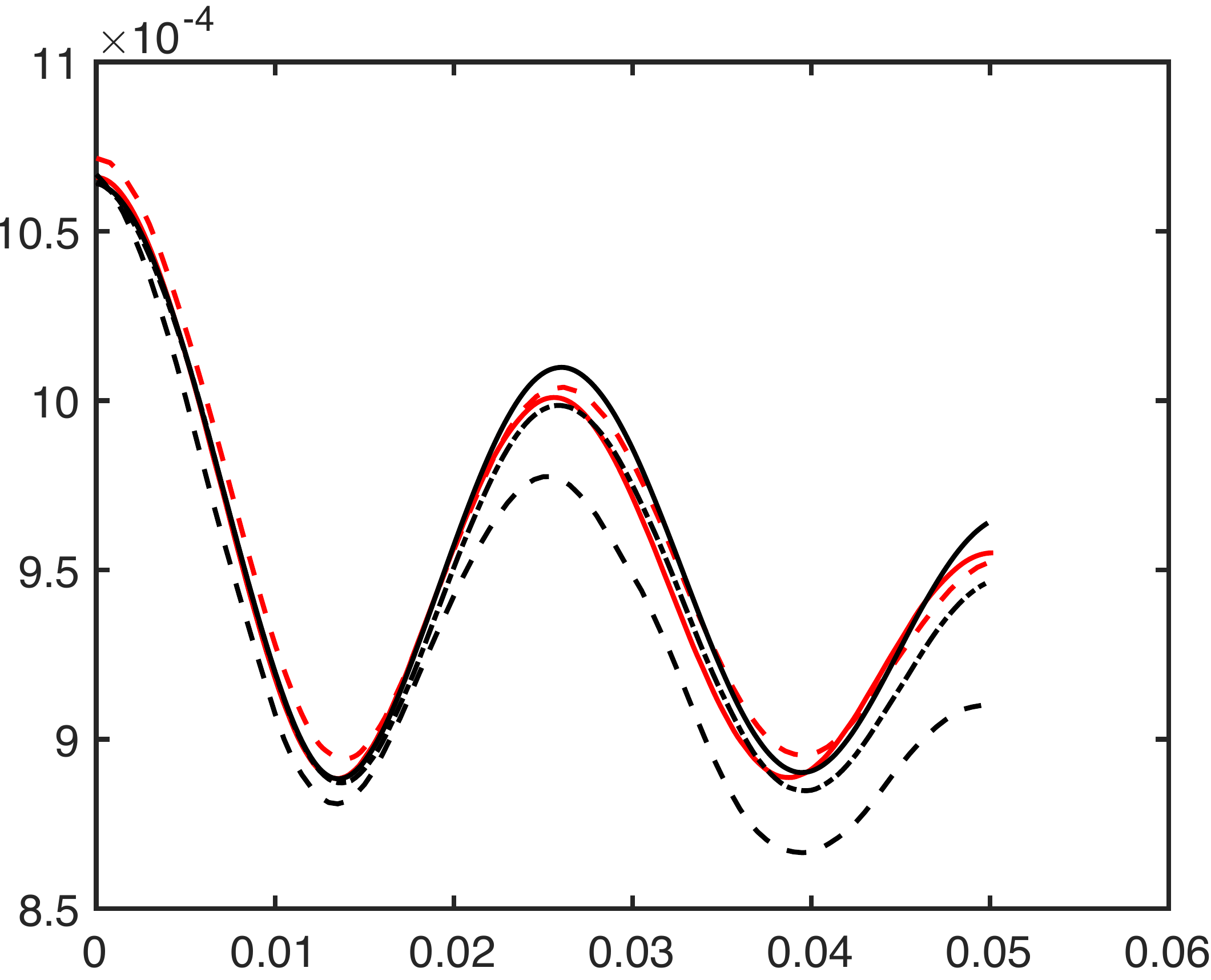}  } }    {\scalebox{0.35}{\includegraphics{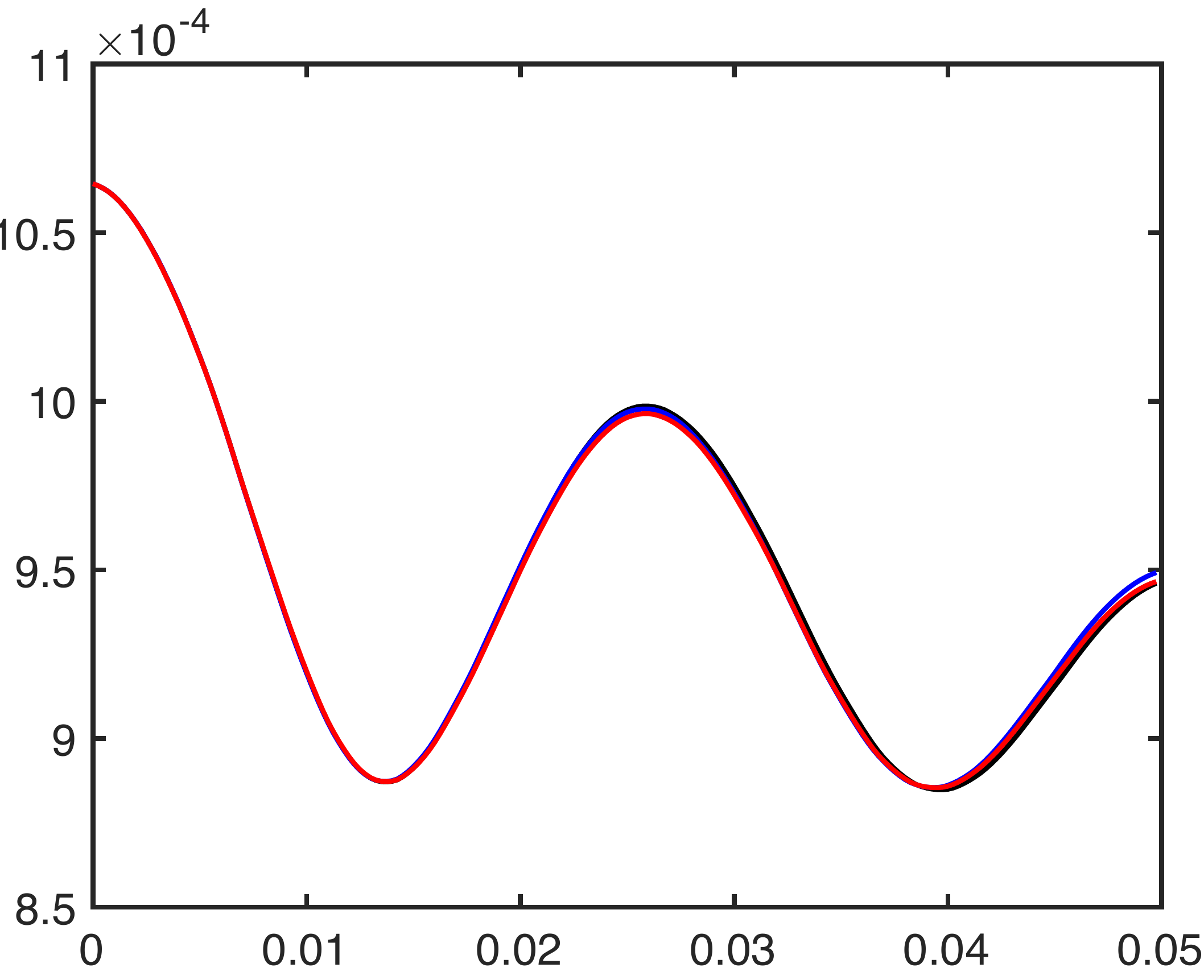}  } }    }
\caption{The moment of the drop versus time. (a) Results computed on a $64^2$ grid (dashed black line), a $128^2$ grid (dot-dashed black line) and $256^2$ grid (solid black line), as well as using a front tracking method (dashed and solid red line).
(b) Results using three different fits computed on a $128^2$ grid.}
 \label{Implemented3}
\end{figure}

A more quantitative comparison is shown if figure \ref{Implemented3}a where the moment of the drop, defined by
\begin{equation}
M(t)=\int \int C(x,y) \bigr( (x-x_0)^2 + (y-y_0)^2 \bigl) dx dy
\end{equation}
is plotted versus time. Here, $(x_0,y_0)$ is the center of the drop. The moment is plotted for for three different grid resolutions ($64^2$, $128^2$ and $256^2$), and it is clear that the results on the two finest grids are relatively close. As a reference we have also plotted results computed using a front-tracking code where the surface tension is found directly from the connected marker particles. The results are shown in red, for two different grid resolutions ($128^2$ and $256^2$ grids). With the exception of the lowest resolution for the machine learning approach all the results are close. In machine learning the fit is not an exact relationship so repeated fitting does not, in general, result in identical results. To test the sensitivity of the results to the specific fit we show the moment versus time in figure \ref{Implemented3}b as found by three different fits. Although a very slight difference can be detected near the end, for most of the time shown all three fits result in essentially identical results.

\section{Conclusions}

Finding a relationship between the volume fractions and the curvature in VOF methods has been a topic of considerable interest since the pioneering paper by \citet{Brackbill:92}. Here we have showed that using machine learning to extract the relationship from a dataset, generated using circles of variable sizes, is able to capture the curvature of a more complex interface with a spatially varying curvature reasonably well. The results suggest that when it is complicated to generate an exact relationships from fundamental considerations, but easy to generate a synthetic dataset from simple examples, this approach may be a viable alternative to more conventional derivations. For the application described here we could keep the learning cases extremely simple, but more complex cases are easily generated. We note that we have not tested the efficiency of this method compared to more traditional ones, and no attempt has been made to make the function efficient.  It is is likely that there are significant opportunities to do so. We also note that while the accuracy is easily established for a given shape and numerical parameters, there is no explicit order that guarantees convergence under grid refinement.

The programs used to generate the data set, do the fitting and generate the neural network, and the flow solver are available as supplementary material.

\section{Acknowledgement}
This research was supported in part by the Consortium for Advanced Simulation of Light Water Reactors, an Energy Innovation Hub for Modeling and Simulation of Nuclear Reactors under U.S. Department of Energy Contract No. DE-AC05-00OR22725.

\newpage
\noindent
{\bf \Large Supplementary Material}
\vspace{25px}

\noindent The codes used for the results reported in the manuscript are included as Supplementary Material. The codes are written in Matlab and require the Neural Network Toolbox. The codes are:

\begin{enumerate}
 \item DataGen.m. A code to generate database (inputs and targets) using circles. The volume fraction of interface cell and its adjacent cells is found as input data and the target is the curvature of interface cell which is constant for a circle. The saved .mat file contains  an input matrix with nine columns and a target matrix with one column. More data could be generated by changing the radius of the circle or the grid spacing.

 \item ML-Train.m. A code to solve an input-output problem with the Neural Network tool box in MATLAB. One may also do the same thing using GUI by typing ``nftool'' in the command window. The code assumes that two variables are defined: input - input data and output - output  data.

 \item VOF-ML.m. A very simple Navier-Stokes solver for a drop in a rectangular box, using a conservative form of the equations.  A 3-order explicit projection method and centered in space discretizations are used. The density is advected by a front tracking scheme and surface tension and variable viscosity is included. The VOF volume fraction is found directly from the front and the surface tension is found by machine learning
 
 \item NNCircle2.m. The neural network simulation function generated by ML-Train.m and called by VOF-ML.m. Generated by Neural Network Toolbox function genFunction, 13-May-2018
 
 \end{enumerate}

\noindent The codes can be downloaded by the following links:
 
 \begin{verbatim} 
 DataGen.m:
 https://www.dropbox.com/s/cmbb221pjdv0g7w/DataGen.m?dl=0
 
 ML-Train.m:
 https://www.dropbox.com/s/1edci5uxnpj0c9t/ML-Train.m?dl=0
 
 VOF-ML.m:
 https://www.dropbox.com/s/zr6ifrxiwgygaww/VOF-ML.m?dl=0
 
 NNCircle2.m
 https://www.dropbox.com/s/pedy9h53w25mkge/NNCircle2.m?dl=0
 
 \end{verbatim}
  
\end{document}